# Specific heat measurement of mesoscopic loops

O. Bourgeois*[1], F. Ong*, S.E. Skipetrov† and J. Chaussy*

*Centre de Recherches sur les Très Basses Températures, CNRS, laboratoire associé à l'UJF et à l'INPG, BP 166, 25 avenue des Martyrs, 38042 Grenoble Cedex 9, France

† Laboratoire de Physique et de Modélisation des Milieux Condensés, CNRS et UJF, BP 166, 25 avenue des Martyrs, 38042 Grenoble Cedex 9, France

**Abstract.** We report highly sensitive specific heat measurement on mesoscopic superconducting loops at low temperature. These mesoscopic systems exhibit thermal properties significantly different from that of the bulk materials. The measurement is performed on a silicon membrane sensor where 450 000 superconducting aluminium loops are deposited through electron beam lithography under an applied magnetic field. Each entry of a vortex is associated to a jump in the specific heat of few thousands of Boltzmann constant $k_B$ indicating the existence of phase transitions. The periodicity of this sequential phase transitions is a nontrivial behaviour and varies strongly as the temperature is decreased. The successive phase transitions are well described by the Ginzburg-Landau theory of superconductivity. The presence of metastable states is responsible for the $n$-$\Phi_0$ ($n$=1, 2, 3…) periodicity of the discontinuities of the measured specific heat.

## INTRODUCTION

Since recent years there has been an increased interest towards the understanding of thermal behavior in nanostructures. For more than two decades, mesoscopic systems have been intensively studied by electronic transport experiments [1]. Although many quantum effects in such systems are now well understood (oscillations of transition temperature in thin superconducting cylinders [2], magnetic flux quantization [3], magnetoresistance oscillations [4], persistent currents [5], etc.), very few experiments shine light on their thermal behaviour [6-7]. Meanwhile, this behaviour is expected to differ significantly from the bulk one, nanometric sample size leading to quantum limitations of thermal transport, new phase transitions and other phenomena specific to mesoscale thermodynamics. Here we report an experimental evidence of nontrivial thermal behaviour of the simplest mesoscopic system – a superconducting loop [7]. By measuring the specific heat of an array of 450 thousands aluminium loops in external magnetic field, we show that they go through a periodic sequence of phase transitions (up to 30 successive transitions) as the magnetic flux $\Phi$ threading each loop is increased. The transitions are separated by $\Delta\Phi = n\Phi_0$ (with $n$ integer and $\Phi_0$ the magnetic flux quantum) and each transition is accompanied by a discontinuity of the specific heat of only several thousands of Boltzmann constants $k_B$. These highly sensitive measurements are made possible by our unique experimental setup that allows us to measure the specific heat with unprecedented accuracy of ~20 femtoJoules per Kelvin, corresponding to energy exchanges as small as few attoJoules (aJ=$10^{-18}$ Joule) at 0.6 K.

## EXPERIMENTAL SET-UP

Our sample is composed of 450 thousands identical aluminium square loops (2 µm in size, $w$ = 230 nm arm width, $d$ = 40 nm thickness, separation of neighbouring loops = 2 µm), patterned by electron beam lithography on a suspended sensor composed of a very thin (4 µm thick) silicon membrane and two integrated transducers: a copper heater and a niobium nitride thermometer [8,7]. The setup is cooled below the critical temperature $T_c$ of the superconducting transition by a $^3$He cryostat and then its specific heat is measured by ac calorimetry. The technique of ac calorimetry consists in supplying ac power to the heater, thus inducing temperature oscillations of the thermally isolated membrane and thermometer (see Ref. 8 for more details). For a carefully chosen operating frequency (in our case, the frequency of the temperature modulation is $f \sim 250$ Hz), the temperature of the system (sensor + sample) follows variations of the supplied power in a quasi-adiabatic way, allowing measurements of the specific heat with a resolution of $\Delta C/C = 5\times10^{-5}$ for signal integration times of the order of one minute. It is then possible to measure variations of the specific heat as small as 1-10 fJ/K (which corresponds to 1-10 thousands of $k_B$ per loop), provided that the specific heat of the sensor (silicon membrane, heater, and thermometer) is reduced to about 10-100 pJ/K.

[1] olivier.bourgeois@grenoble.cnrs.fr

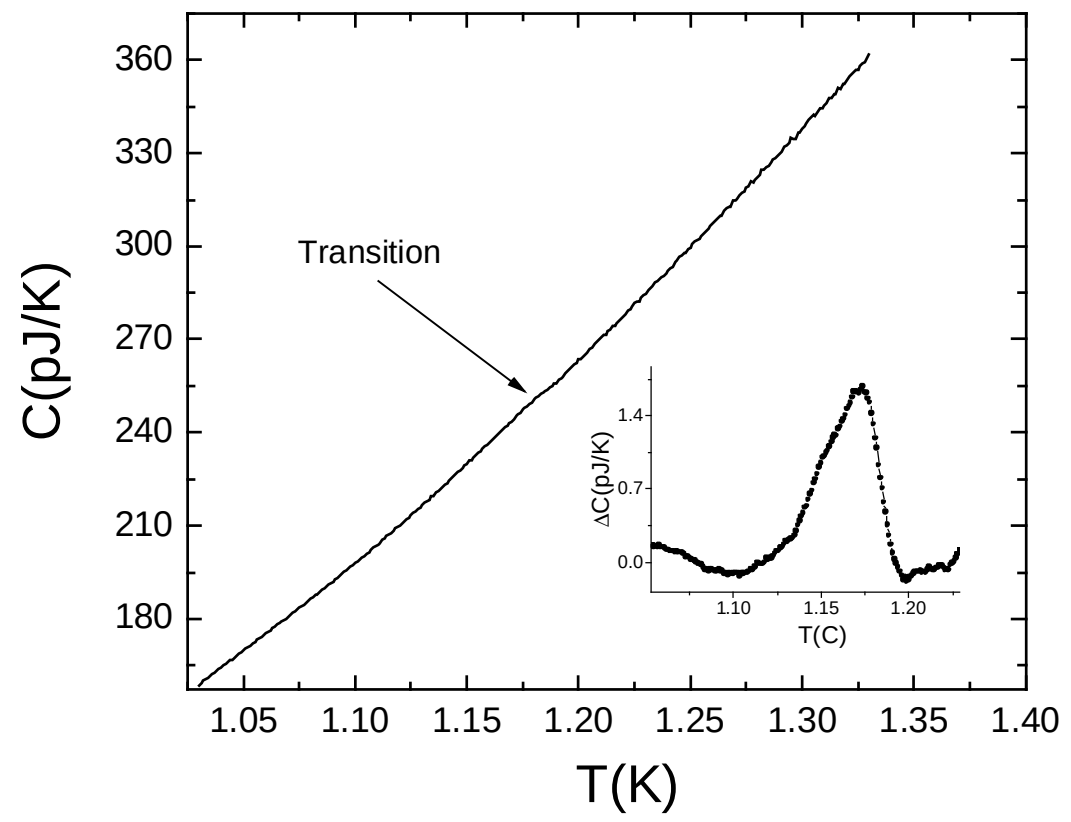

**FIGURE 1.** Heat capacity of the total system (sensor and aluminium). This illustrates the small amplitude of the specific heat jump at the superconducting phase transition. In the inset the background was subtracted to give an order of magnitude of the jump at the transition temperature of 1.2K. The temperature modulation used for the measurement ranges between 1 and 10mK.

We emphasize that working with lithographied thin film systems implies that the mass of the sample is very small, usually few tens of nanograms, leading to major difficulties when one wants to measure thermal signatures of quantum phenomena in such systems. Hence, the above extreme sensitivity is absolutely necessary. Since the temperature oscillates with typical amplitudes of few mK, our experimental apparatus can detect energy exchanges of only few aJ at the lowest temperature of 0.6 K used in our experiments.

In the absence of magnetic field, the transition of the sample into the superconducting state is observed around $T_c$ = 1.2 K with the total specific heat discontinuity of 2 pJ/K as it can be seen in the Fig. 1. In terms of jump in specific heat at the superconducting transition, this corresponds to much less than what is expected for 80 ng of aluminium (4pJ/K) deposited on the sensor. It can be explained either by the fact that part of the aluminium deposited is oxidized but also by a hard to implement background subtraction. It gives anyhow an idea of the difficulties of measuring the specific heat signature arising at the superconducting second order phase transition of a low mass mesoscopic system.

## RESULTS

The mesoscopic thermodynamic behaviour of the superconducting loops is studied through the measurement of the specific heat under an applied magnetic field. The specific heat can be defined as the second derivative of the Ginzburg-Landau free energy $F_{GL}$:

$$C(H,T) = -T\frac{\partial^2 F_{GL}(H,T)}{\partial T^2} \qquad (1)$$

The free energy is given by :

$$F_{GL} = \alpha|\psi|^2 + \frac{\beta}{2}|\psi|^4 + \frac{1}{4m}\left|\left(\frac{\hbar}{i}\vec{\nabla} - \frac{2e}{c}\vec{A}\right)\psi\right|^2 \qquad (2)$$

Then the only contribution for the flux dependent specific heat comes from the third term; the temperature dependence coming from the order parameter $|\Psi|^2$ and the magnetic field contribution is taken into account in the vector potential **A**.

If we first focus our attention to the temperature very close to the transition temperature, we do not expect major signature in *C(H)*. Indeed, the coefficient $\alpha$ is linear with temperature and $\beta$ can be considered as a constant and because $\Psi = -\alpha/\beta$ then no signal versus magnetic field is expected in the specific heat. This is related to the fact that the critical magnetic field is linear near $T_c$. One has to go far from $T_c$ and take into account the temperature dependence of $\beta$ in the expression of the free energy to get nontrivial variation of the specific heat signal versus the magnetic field threading the loops. The fluxoid is quantized in such geometry then we expect that the total specific heat signal shows $\Phi_0$ periodic oscillations versus the magnetic flux at the thermodynamic equilibrium. At a given magnetic flux $\Phi$ threading the loop, we assume that the superconducting order parameter takes the form $\Psi = f(\rho)e^{in\phi}$, where the vorticity $n$ is a number of "giant" magnetic vortices in the loop and we use cylindrical coordinates with the $z$ axis perpendicular to the plane of the loop.

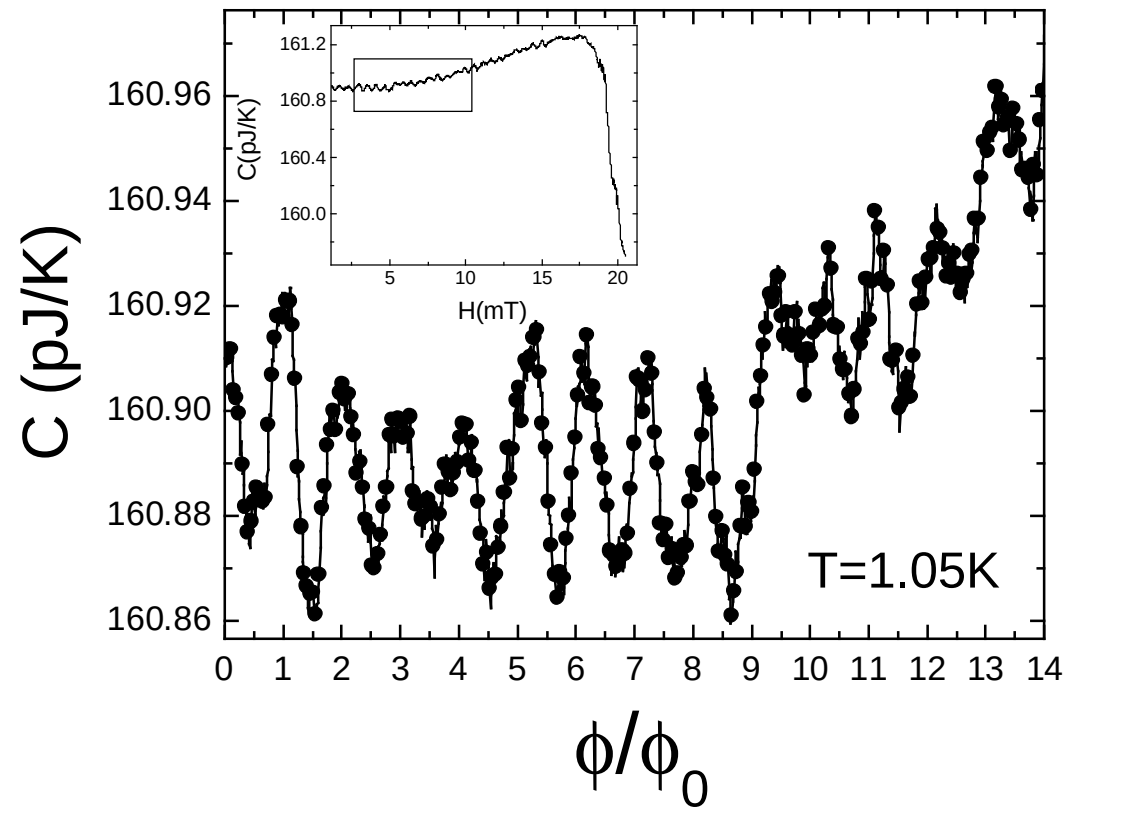

**FIGURE 2.** Heat capacity of the mesoscopic superconducting loops versus normalized magnetic flow ($\Phi_0$ corresponds to a magnetic field of 0.58mT in a loop of 1.8μm in diameter) performed at 1.05K. The specific heat exhibits $\Phi_0$ periodic oscillations. In the inset we present the complete variation of specific heat under magnetic field where the critical magnetic field is observed around 19mT. The rectangle represents the part of the curve enlarged in the main frame of the figure.

In Fig. 2 we present the specific heat measurement under magnetic field at 1.05K. The major jump at 20mT represent the superconducting to normal state phase transition at the critical magnetic field of the aluminium loops (see the inset). Meanwhile, significant oscillations appear at low field with a clear periodicity of 0.58mT which corresponds to the required magnetic field to create one giant vortex (one quantum flux) in a loop of diameter 1.8μm. These oscillations, as expected, are the signature of the penetration of vortices inside the loops. Each giant vortex entry is associated to a mesoscopic phase transition and hence a specific heat jump corresponding to $5 \times 10^{-14}$J/K, i.e. $9 \times 10^{-20}$J/K per loop (7500 $k_B$) is observed. The existence of such kind of phase transitions was predicted for holes in a superconductor or for disks in the framework of the Ginzburg-Landau (GL) theory of superconductivity, which can also be applied to superconducting loops [9]. These measurements are in a perfect agreement with previous measurement by susceptibility measurements [10] micro-Hall probe magnetometry [11,12] or by more recent electrical measurements [13] on mesoscopic superconducting disks or rings.

At lower temperature, we observed multiple phase transitions also but between states having a vorticity larger than one. Around 0.8 K, the specific heat is $2\Phi_0$ periodic and at 0.6 K as it is shown in Fig. 3 the signal becomes $3\Phi_0$ periodic. These mesoscopic phase transitions involving a high number of vortices are characterized by a larger specific heat jump as well as a larger magnetization change [14].

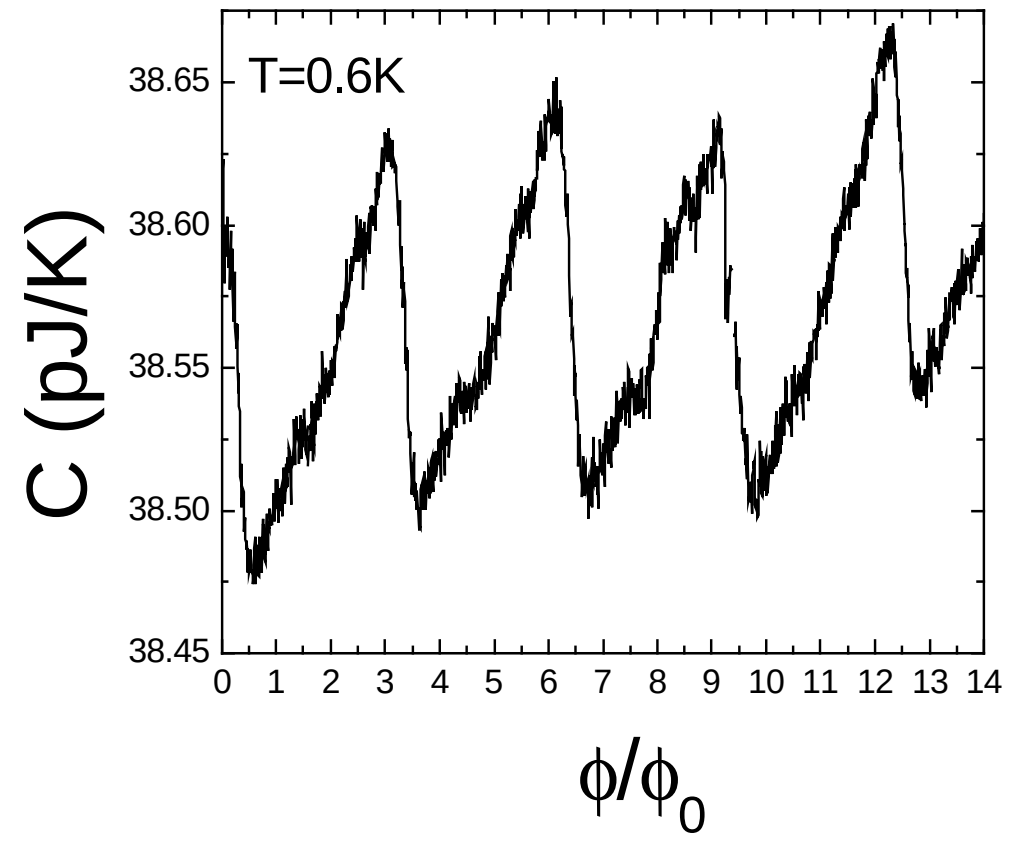

**FIGURE 3.** Heat capacity of the superconducting loops measured at 0.6K (same normalization for the x axis as in Fig. 2). At low temperature, $3\Phi_0$ periodic specific heat jumps are observed with amplitude of $15 \times 10^{-14}$J/K.

This specific behaviour of multiple flux jumps or flux avalanches, signatures of simultaneous entrance of several giant vortices in the loops, cannot be explained by equilibrium thermodynamics. If we calculate the position of the specific heat jump in the equilibrium case, the periodicity corresponding to the cross of two free energy parabola is never greater than one $\Phi_0$. The measurement shown in Fig. 3 indicates that the system actually explores metastable states, staying in a local minimum but not in the stable state. The environmental noise (thermal or electromagnetic) is not enough to overcome the energy barrier between the metastable states and the stable states. The GL characteristic time (which can be greater than $10^7$ seconds) of that type of event is much larger than the current time of an experiment. This means that the system stays in the out of equilibrium state until it reaches an unstable state.

This behaviour has been perfectly described in the framework of the GL theory through numerical calculations. By minimizing the free energy and founding the correct wave function we were able to calculate the specific heat of the system. We assume a circular loop of the same average perimeter as the actual square loop and the value of the GL coherence length at zero temperature $\xi(0)$ = 130 nm, evaluated from transport measurements. The correct free energy was found by finding the wave function minimizing the free energy through numerical computations. As in the experimental data, several values of vorticity are possible at a given value of $\Phi$. In the thermodynamic equilibrium, the transitions between the states with vorticities $n$ and $n + 1$ (so-called "giant vortex states") occur at $\Phi = (n + 1/2)\ \Phi_0$, which minimizes the free energy and makes the specific heat to follow the lowest of curves corresponding to different $n$, as it can be seen in the Fig. 4. This results in oscillations of $C(\Phi)$ with a period of $\Phi_0$, in agreement with the experiment. The amplitude of oscillations decreases when $T$ approaches $T_c$, and hence the oscillations are likely to be masked by noise at $T$ close to $T_c$, which explains why we did not succeed to observe experimentally any fine structure in $C$ at $T > 0.93\ T_c$.

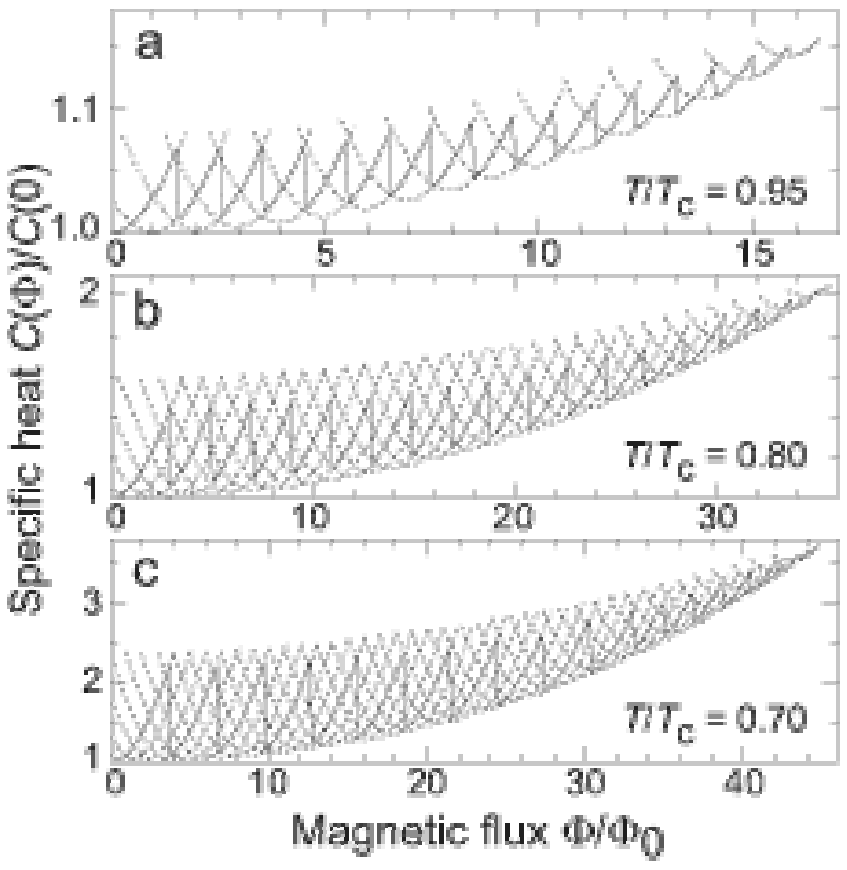

**FIGURE 4.** Normalized specific heat versus the magnetic flux of a superconducting loop computed from the numerical solution of the GL equation for three different temperatures. Dashed lines correspond to different states with constant vorticity (0, 1, 2 and 3 etc…).

The jump in the specific is found to appear when the state becomes completely unstable. This arises when the circulating current reaches the critical current of the superconducting loop. The numerical calculation gives a rather good qualitative description of what was experimentally observed. A change of periodicity along with the magnetic field is observed in both. In Fig. 3, the signal changes from a 3 $\Phi_0$ periodic signal to a one $\Phi_0$ signal. Same king of change are also observed in the Fig. 4.

Finally, the experimental demonstration of oscillatory behaviour and jumps of the specific heat at vortex entrance in a superconducting loop reveals the great interest of studying thermodynamic behaviour of mesoscopic systems (mesoscale thermodynamics). In fact, the possibility of measuring specific heat of nanoscale objects with very high accuracy opens quite interesting prospects. For instance, at low temperatures great opportunities exist for studying phase transitions between states with the same number but different configurations of vortices in mesoscopic superconductors, the superconducting phase transition in nanometer sized grains [15], thermal signatures of phase coherence in normal metals or quantum heat transfer in single nanocrystals or nanowires [16].

## ACKNOWLEDGMENTS

The authors want to thank the Groupe de Biothermique et de Nanocalorimétrie (E. André, P. Lachkar, and J-L. Garden), and P. Gandit, P. Brosse-Maron, J-L. Bret for useful help and assistance in cryogenics and electronics and fruitfull discussion with R. Maynard, G. Deutscher and H. Pothier.